\documentclass[twocolumn,8pt]{asme2e}
\usepackage[utf8]{inputenc}
\usepackage{color,soul}
\usepackage{xcolor}
\usepackage{listings}
\usepackage{graphicx}
\usepackage{float}
\usepackage{mathtools}
\usepackage{appendix}
\usepackage{geometry}
\usepackage{hyperref}

\hypersetup{
    colorlinks=true,
    linkcolor=magenta,
    filecolor=magenta,  
    urlcolor=magenta,
    citecolor=magenta,
}

\definecolor{codegreen}{rgb}{0,0.6,0}
\definecolor{codegray}{rgb}{0.5,0.5,0.5}
\definecolor{codepurple}{rgb}{0.58,0,0.82}
\definecolor{backcolour}{rgb}{0.95,0.95,0.92}

\lstdefinestyle{mycodestyle}{
  backgroundcolor=\color{backcolour},
  commentstyle=\color{codegreen},
  keywordstyle=\color{magenta},
  numberstyle=\fontsize{6}{11}\color{codegray},
  stringstyle=\color{codepurple},
  basicstyle=\fontsize{7}{11}\ttfamily,
  breakatwhitespace=false,
  breaklines=true,
  captionpos=b,
  keepspaces=true,
  numbers=left,
  numbersep=2pt,
  showspaces=false,
  showstringspaces=false,
  showtabs=false,
  tabsize=2
}
\usepackage{graphicx}
\graphicspath{ {./Figures/} }

\usepackage{mathrsfs}
\usepackage{mathtools}
\usepackage{phoenician}
\newcommand{\pp}{\;\mathclap{\reflectbox{$\mathrm{P}$}}\mathrm{P}}

\newcommand{\rev}[1]{ #1 }
\newcommand{\ttsml}[1]{{\small \texttt{#1}}}

\usepackage{float}

\title{SIMD-Optimized Search Over Sorted Data}

\author{Benjamin Mastripolito\thanks{Address all correspondence to this author.} \\
    \affiliation{X Computational Physics Division\\
	Los Alamos National Laboratory\\
	Los Alamos, New Mexico, 87545\\
	bmastripolito@lanl.gov
    }
}

\author{Nicholas Koskelo\\
    \affiliation{
	Harvey Mudd College and \\
	Los Alamos National Laboratory\\
	Los Alamos, New Mexico, 87545\\
    nkoskelo@lanl.gov
    }
}

\author{Dylan Weatherred \\
    \affiliation{
    University of New Mexico and\\
	Los Alamos National Laboratory\\
	Los Alamos, New Mexico, 87545\\
	dweatherred@lanl.gov
    }
}

\author{David A. Pimentel\thanks{Address all correspondence to this author.} \\
    \affiliation{X Computational Physics Division\\
	Los Alamos National Laboratory\\
	Los Alamos, New Mexico, 87545\\
	davidp@lanl.gov
    }
}
\author{Daniel Sheppard \\
    \affiliation{X Computational Physics Division\\
	Los Alamos National Laboratory\\
	Los Alamos, New Mexico, 87545\\
	danielsheppard@lanl.gov
    }
}
\author{Anna Pietarila Graham \\
    \affiliation{High Performance Computing Division\\
	Los Alamos National Laboratory\\
	Los Alamos, New Mexico, 87545\\
	annap@lanl.gov
    }
}

\author{Laura Monroe \\
    \affiliation{
    High Performance Computing Division\\
	Los Alamos National Laboratory\\
	Los Alamos, New Mexico 87545\\
	lmonroe@lanl.gov
    }
}

\author{Robert Robey \\
    \affiliation{X Computational Physics Division\\
	Los Alamos National Laboratory\\
	Los Alamos, New Mexico 87545\\
	brobey@lanl.gov
    }
}

\begin{document}
\maketitle

\begin{abstract}
{\it  Applications often require a fast, single-threaded search algorithm over sorted data, typical in table-lookup operations. We explore various search algorithms for a large number of search candidates over a relatively small array of logarithmically-distributed sorted data. These include an innovative hash-based search that takes advantage of floating point representation to bin data by the exponent. Algorithms that can be optimized to take advantage of SIMD vector instructions are of particular interest. We then conduct a case study applying our results and analyzing algorithmic performance with the EOSPAC package. EOSPAC is a table look-up library for manipulation and interpolation of SESAME equation-of-state data. Our investigation results in a couple of algorithms with better performance with a best case 8$\times$ speedup over the original EOSPAC Hunt-and-Locate implementation. Our techniques \rev{are generalizable} to other instances of search algorithms seeking to get a performance boost from vectorization.}
\end{abstract}

\section{Introduction}
\rev{Fast search over a} sorted array is a common problem in computer science. \rev{Many algorithms developed for this purpose have been optimized for serial processing with a simple (single) processor model.}

\rev{These algorithms often cannot be compiled to make use of specialized SIMD instructions (a process called vectorization)}. For example, the hunt-and-locate algorithm uses each index found as the starting index for the subsequent search iteration. This loop-carried dependency prevents  vectorization. In conventional algorithm analysis, algorithms are often judged by their asymptotic behavior. However, in real-life applications the constants in the scaling behavior and suitability for vectorization matter, \rev{and are not necessarily addressed by an asymptotic analysis.}

\rev{In this paper we conduct a case study of new and existing algorithms adapted for vectorization which find the index of the largest element of an array less than a given target value.} We will implement and test the algorithms within an equation of state library, EOSPAC~\cite{info:lanl-repo/lareport/LA-UR-18-31814}, which is employed by several large physics codes to interpolate equation of state and other material property data. The searched data are density and temperature arrays of a particular SESAME~\cite{info:lanl-repo/lareport/LA-UR-92-3407} material that is loaded and used by EOSPAC. \rev{EOSPAC provides interpolation mechanisms, which involve searching for values nearest to a sequence of targets within a small table. Applications which utilize this routine often pass a much larger number of target values than there are values in the table. They also often use all threads of the host machine, so parallelism through multi-threading is not an option for us. Due to these constraints, and the fact that user applications typically run on SIMD-enabled architectures, we determined that a more optimized approach was possible. We will investigate how the algorithms' performance depends on array size, cache usage, portability, and other factors.}

\section{Background}

We first quickly review the applicable historical background of search algorithms. Then we give a short overview of the application, EOSPAC, that we used in this study.

\subsection{Search Algorithms}
Search algorithms are some of the oldest algorithms in computer science, dating back to the early days of digital computers. In this paper, we are concerned with searching in an array which is already sorted. Knuth gives an overview of searching ordered data in \textit{The Art of Programming, Volume 3: Sorting and Searching}~\cite{knuth1998art}. We briefly discuss here in general terms the search algorithms we compared.

The simplest form of search is a linear search, where every element of the array is compared against the target value. \rev{Linear search provides a good lower bound for search efficiency.}

Binary (bisection) search is an improvement on linear search that takes advantage of a sorted array. Knuth~\cite{knuth1998art} credits John Mauchly in 1946 for the earliest reference to binary searching~\cite{mauchly1946theory}. The mid-point of the array is selected, a comparison with the target value is made, then the search is narrowed by half: either the right or left half of the array is selected for the next iteration. The division of the search space in half at each step results in $\mathcal{O}(log (n))$ operations for every search, where $n$ is the number of elements in the data array.

The hunt-and-locate~\cite{press1992numerical} search algorithm utilizes the binary search technique in its second of two steps. In the first step, an exponential jumping pattern is used to bound the target value within the array. It then performs a binary search within these bounds, locating the target. If the search is successful, these bounds are used a starting point for the next search. If the next target is near or  within these bounds, the next search will be faster than simply a binary search.

The skip-list search algorithm~\cite{skiplist89} further builds on the idea of intelligently partitioning the array to decrease the number of operations needed to locate the target value. The algorithm involves creating a "skip-list", which is a reduced copy of the array including only every $i$th element, \textit{skipping} the others. The skip-list can be thought of as linear array version of a search tree, but with potentially better cache performance. When a search is performed, the skip-list is searched over with a binary search, providing the bounds with which to search between for the target in the array. Multiple skip-lists can be combined in this way to further improve search times.

Finally, for some arrays, a spatial hash or binning approach can perform very well. The values in the array are separated by a hash function into "bins", partitioning the array. The same hash function is then used on the target, indicating which bin must be searched. An implementation involves analyzing the array to build an analytic hash function as demonstrated in Robey and Nicholaeff, et al.~\cite{robey2013hash}. Since a hash function does not need a comparison operation, it can approach constant-time performance.

\subsection{Domain Problem: Accessing tabular SESAME data via EOSPAC}
The EOSPAC package~\cite{info:lanl-repo/lareport/LA-UR-18-31814} is a collection of C routines that can be used to access the SESAME data library~\cite{info:lanl-repo/lareport/LA-UR-92-3407}, which contains tabulated data representing various equation of state (EOS) properties for a diverse collection of materials spanning more than fifty years of experimentation and modeling. 

An equation of state is a formula describing the interconnections between various macroscopically measurable properties of a system. In physics and thermodynamics, an EOS is a constitutive equation describing the state of matter under a given set of physical conditions. It provides a mathematical relationship between two or more of that matter's state functions, such as its temperature, pressure, volume, or internal energy. Equations of state are useful in describing properties of fluids, mixtures of fluids, and solids.

The EOSPAC package is designed to be used by physics codes that may be written in multiple languages and on multiple platforms. EOSPAC has a long history with its Fortran predecessors dating back forty years.

EOSPAC accesses and loads SESAME data tables and provides interpolation mechanisms for use by a host application.

We chose the EOSPAC software package as an exemplar for our analysis of search algorithms for several reasons:
\begin{itemize}
    \item - EOSPAC provides a mature software and development framework that is dependent upon tabular data search
    \item - Many other software packages are dependent upon EOSPAC, and require it to be as efficient as possible on various high performance computing (HPC) systems
    \item - The SESAME tables are relatively small, and they contain sorted, irregularly-spaced data grids that must be efficiently searched to bound arbitrarily-spaced and arbitrarily-large lists of interpolator input values.
    \item - Large numbers (i.e., millions/billions) of searches typically create an execution bottleneck that may degrade code performance.
\end{itemize}
Historically, EOSPAC has utilized the Numerical Recipes \cite{press1992numerical} \it{Hunt-and-Locate} \rm{algorithm} described above. Since EOSPAC has historically been a serial software package, the selected search algorithm has provided sufficient performance for supported platforms and host applications. With the emergence of vectorized processors and threading on GPUs, the current hunt-and-locate serial search is no longer appropriate because of the loop-carried dependency.

\section{Methodology}
In this section we discuss how we apply concepts from Robey and Zamora~\cite{robey_zamora} to search algorithms with the goal of better taking advantage of vectorization. Each algorithm takes a \textit{targets array} ($Y$, of size $m$), and a \textit{data array} ($X$, of size $n$). They perform a lower-bounds search, returning an array of size $m$ containing the index of the last element of $X$ which is less than or equal to each element of $Y$. If $X$ contains no such lower bound for a given $y \in Y$, the index $0$ is chosen. In the case that the element is higher than the highest of \rev{$X$}, the index $n - 1$ is chosen.

The data array ($X$) is monotonically increasing, approximately logarithmically distributed, and much smaller than the search array ($Y$). For our tests, $m=5000000$, and $n=110$. The way in which the algorithms are vectorized take advantage of these data characteristics.~\cite{press1992numerical} Using OpenMP~\cite{openmp} SIMD pragmas, we configure the outer loop over $Y$ to be the vectorized loop. This is optimal because the outer loop runs for many more iterations than the inner loop over $X$. The SIMD pragmas were added to the OpenMP 4.0 standard released in 2013 to provide a more portable syntax than the ad-hoc directives provided by some vendors.

A common optimization we utilized was the removal of branching. Compilers, when performing vectorization, will sometimes fail to vectorize when there is complex branching. We removed branching in some places and left it in others because sometimes the compiler succeeds in vectorizing the branch statements, using masking. The masking that occurs we found to be generally faster than removing branching manually. When we were able to remove branching, we used a "branchless choice" inlined function to replace if/else statements:

\begin{lstlisting}[caption={Branchless choice function: returns vTrue if {\small \texttt{c}} is not 0, vFalse otherwise. Arguments must be integers.},label={lst:bchoice}]
bchoice(c, vTrue, vFalse)
    return (c * vTrue) | (!c * vFalse)
\end{lstlisting}

This sort of technique is also used for branchless min and max functions. It only works for integers, as it makes use of a bitwise operation.

\rev{We have made additional, minor optimizations in the C code [\ref{appendix}], such as storing the minimum and maximum of the data array in constant variables outside the loop, where needed. These minor optimizations are not included in the following pseudocode for the purpose of clarity, and are not as impactful as the major optimizations, such as the removal of branching.}

\subsection{Hunt and Locate}
\label{sect:hunt_and_locate}
The \textit{Hunt and Locate} algorithm is based upon the algorithm described by Press, Teukolsky et al. in Numerical Recipes~\cite{press1992numerical}. We chose to measure the performance of our other algorithms relative to \textit{Hunt and Locate} because it is the original search algorithm implemented in EOSPAC~\cite{info:lanl-repo/lareport/LA-UR-18-31814}. 

\begin{lstlisting}[caption={Modified Hunt and Locate},label={lst:hunt_n_locate}]
hunt_n_locate(X, n, Y, m)
    lowbounds[m]
    for(int i = 0; i < m; i++):
        if Y[i] < X[0]:
            lowbounds[i] = 0
        else if Y[i] > X[n-1]:
            lowbounds[i] = n-1
        else:
            j = 1
            start = 0
            end = 1
            while end < n && Y[i] > X[end]:
                start = end
                end <<= 1
            // Binary search within bounds
            end = min(end, n-1)
            start = binary_search(
                max(start, 0), min(end, n-1),
                x, target)
            lowbounds[i] = start
    return lowbounds
\end{lstlisting}

We modified \textit{Hunt and Locate} to take advantage of vectorization \rev{by removing the saving of search bounds between each iteration of the loop over the target array [Listing~\ref{lst:hunt_n_locate}]. By simplifying the logic in this way as well as removing branching for bounds checking, an increase in performance is possible through vectorization. Also, since the search array is not ordered, there is little advantage to maintaining search bounds between iterations anyways.}

\subsection{Branchless Binary Search}
\label{sect:binary_search}
\rev{ The following is a simple binary search where branching has been removed from search bound selection (lines 10 and 11) and bounds checking (13 and 14) using the branchless choice function [Listing~\ref{lst:bchoice}]. As we will discuss later, this was done to improve performance when SIMD is enabled during compilation. However, it still contains a loop of a length which is unknown at compile time, which makes it a non-ideal candidate for SIMD optimization. }

\begin{lstlisting}[caption={Branchless binary search},label={lst:binary_search}]
binary_search(X, n, Y, m)
    lowbounds[m]
    for(int i = 0; i < m; i++):
        lower = 0
        upper = n
        // Search
        while((upper - lower) > 1):
            mid = (lower + upper) / 2
            condition = Y[i] < X[mid]
            upper = bchoice(condition, mid, upper)
            lower = bchoice(!condition, mid, lower)
        // Bounds checking
        lower = bchoice(Y[i] < X[0], 0, lower)
        lower = bchoice(Y[i] > X[n-1], n-1, lower)
        lowbounds[i] = lower
    return lowbounds
\end{lstlisting}

\subsection{Skiplist Search}
\label{sect:skiplist_search}
\textit{Skiplist Search} generates a table of indices \rev{(\ttsml{skplst}) into the data array. We chose to build this table so that each index corresponded to its own cache line. When searching for a target, first, a binary search is performed over the table. Then, the resultant value is used to bound a subsequent linear search.} The only improvement made to \textit{Skiplist Search} for vectorization was to eliminate branching in bounds checking.

\begin{lstlisting}[caption={A skiplist implementation of a binary search},label={lst:skiplist}]
skiplist_search(X, n, Y, m)
    skplst[n/8], lowbounds[m]
    for(i = 0; i < n/8; i++):
        skplst[i] = X[8 * i]
    for(i = 0; i < m; i++):
        j = binary_search(skplst, Y[i])
        upper = n
        if j < n/8-1:
            upper = (j+1)*8
        lower = linear_search(X, 8*j, upper, Y[i])
        // Bounds checking
        lower = bchoice(Y[i] < X[0], 0, lower)
        lower = bchoice(Y[i] > X[n-1], n-1, lower)
        lowbounds[i] = lower
    return lowbounds
\end{lstlisting}

Note that the setup cost for \textit{Skiplist Search} is $O(\frac{n}{8})$ in time and $n + n/8$ in storage. This is larger than the setup cost for binary search. 


\subsection{Logarithm Hash Search}
\label{sect:log_hash_search}

\textit{Logarithm Hash Search} uses a hash table $H$, the entries of which are indices into the data array, $X$. Each hash table entry $H_i$ contains the smallest index $j$ in the set $\{j\textrm{ | }\textrm{hash}(X_j)=i\}$ where $X_j$ is the $j$th element of the data array $X$. To perform a search for $Y_k$ in $X$, the hash of the target value $v=\textrm{hash}(Y_k)$ is used to index into the hash table at $H_v$. Then a binary search is performed between $H_v$ and $H_{v+1}$. The handling of boundary conditions is shown on lines 24 and 25 of Listing~\ref{lst:hashsearch}.

The hash function $\textrm{hash}(x)$ is essentially just $log_b(x)$ where $b$ is calculated such that the hash function fits every value of the data array $X$ within $H$ of desired size $|H|$. In other words, any value we search for that is within $[X_{min},X_{max}]$ must satisfy the condition $0 \leq \textrm{hash}(x) < |H|$. We calculate the base $b$ with Equation~\ref{eq:hashfuntionbase10}.

\begin{equation}
    b = 10^{
        \frac{\textrm{log}_{10}(X_{max})
            - \textrm{log}_{10}(X_{min})}
            {|H|}
    }
    \label{eq:hashfuntionbase10}
\end{equation}

The hash function also needs an offset to ensure that $0 \leq \textrm{hash}(x) < |H|$. Equation~\ref{eq:hashoffset} shows the equation for the offset.

\begin{equation}
   \textrm{hash}(x) = \lfloor \textrm{log}_b(x) \rfloor - \lfloor \textrm{log}_b(X_{min}) \rfloor
   \label{eq:hashoffset}
\end{equation}

To compute the value of $H_i$, we simply take the minimum index of all data values $X_j$ where $\textrm{hash}(X_j) = i$. This results in Equation~\ref{eq:hashentrycalc}.

\begin{equation}
   H_i = \textrm{min}_j(\textrm{hash}(X_j) = i )
   \label{eq:hashentrycalc}
\end{equation}

If there is no $X_j$ where $\textrm{hash}(X_j) = i$, $H_i = H_{i+1} - 1$, or $n-1$, if $i = n-1$.

\begin{lstlisting}[caption={A hash-based search approach},label={lst:hashsearch}]
hash(x, base, offset)
    return log(x) / log(base) + offset
    
log_hash_search(X, n, Y, m)
    lowbounds[n], H[H_size]
    base = pow(10, (log(X[m-1]) - log(X[0])) / H_size)
    offset = floor(log(X[0]) / log(base))
    for(i = 0; i < H_size; i++):
        H[i] = m-1
    for(i = 0; i < m; i++):
        index = hash(X[i], base, offset)
        H[index] = min(H[index], i)
    for(i = 0; i < H_size; i++):
        upper = H[i]
        while H[i-1] > upper:
            H[i-1] = upper - 1
            i -= 1
    for(i = 0; i < n; i++):
        j = hash(Y[i], base, offset)
        start = max(0, H[j])
        end = min(n-1, H[min(j+1, H_size-1)])
        lower = linear_search(x, start, end, Y[i])
        // Bounds checking
        lower = bchoice(Y[i] < X[0], 0, lower)
        lower = bchoice(Y[i] > X[n-1], n-1, lower)
        lowbounds[i] = lower
    return lowbounds
\end{lstlisting}

In Listing \ref{lst:hashsearch}, linear search is used to search between bounds calculated using the hash table. The arguments are: \ttsml{Array, start\_index, end\_index, searchFor}. We tested hash search with both linear and binary search as the interior search algorithm. Linear search generally performs better with vectorization as each lane has equally divided work and a more predictable control flow. However, this is only true when the index distance between hash table entries is small. Choosing a large value for \ttsml{H\_size}  ($|H|$) typically ensures this if the data is somewhat logarithmically distributed.

\subsection{Exponent Hash Search}
\label{sect:exponent_hash_search}
To better take advantage of vectorization, we also explored a hash search that extracts the base 2 exponent from the exponent bits of double-precision IEEE 754 floating point numbers. Aside from the hash function itself, \textit{Exponent Hash Search} is the same as \textit{Logarithm Hash Search}. The trade-off is between increased vectorized performance and potentially longer linear searches. \rev{With \textit{Exponent Hash Search}, the number of data array accesses increases with the number of data array values which share a base 2 exponent. The more data array elements which bin into a power of 2, the longer the search range is for an element in that bin. In contrast, the base used for binning in \textit{Logarithm Hash Search} is unconstrained, so in some cases it can perform better.}

\section{Methods}

We evaluated the performance and portability of the different search algorithms with and without vectorization on the Darwin computing cluster at Los Alamos National Laboratory~\cite{darwin_cluster}. 

The Darwin system is a heterogeneous cluster designed to be a test bed for new technologies and architectures. For this study we included the following CPU models: Intel Skylake Gold\footnote{Linux; 3.10.0-1127.19.1.el7.x86\_64; \#1 SMP Tue Aug 25 17:23:54 UTC 2020; x86\_64; GNU/Linux; \hl{avx512f}}, Intel KNL Xeon Phi\footnote{Linux; 3.10.0-1127.19.1.el7.x86\_64; \#1 SMP Tue Aug 25 17:23:54 UTC 2020; x86\_64; GNU/Linux; \hl{avx512pf}}, and AMD EPYC\footnote{Linux; 3.10.0-1127.19.1.el7.x86\_64; \#1 SMP Tue Aug 25 17:23:54 UTC 2020; x86\_64; GNU/Linux; \hl{avx2}}. 

We used the gcc/9.3.0 and Intel/20.0.2 compilers. The compiler flags (listed in appendix \ref{appendix}) are those recommended by Robey and Zamora~\cite{robey_zamora} to get optimal vectorization while maintaining numerical correctness as close to the IEEE floating point standard~\cite{IEEE-standard}. All the tests were run on a single core.

The table used as the search array is the density axis of the SESAME~\cite{info:lanl-repo/lareport/LA-UR-92-3407} table 301 for material 3720 (distributed with the release of EOSPAC~\cite{info:lanl-repo/lareport/LA-UR-18-31814}). There are 111 density values ranging from 0 g/cc to 54000 g/cc. We used a batch size ($m$) of 5 million, and averaged over 20 trials. The target values ($Y$) were generated randomly from a uniform distribution ranging from $10^{-10}$ to $10^{10}$. We also experimented with a Gaussian distribution of target values but it had no effect on the results.

To study the cache performance of the algorithms we used Likwid~\cite{likwid} which employs hardware counters for the purpose of performance analysis. In particular, we measured the L2 cache bandwidth, L2 cache miss rate, L3 cache bandwidth, and L3 cache miss rate. The Likwid measurements were made on a Intel Skylake Gold node.

It is desirable for a production library to have a single algorithm that gives good performance across multiple architectures (or at least as few algorithms as possible). For a quantitative measure of each algorithm's portability and performance across the different CPU models and compilers, we used the Pennycook method~\cite{DBLP:journals/corr/PennycookSL16}. The metric is computed using Equation~\ref{eq:pennycook}.

\begin{equation}
    \pp(a,p,C) = \begin{dcases}
		\frac{|C|}{{\sum_{\textit{i} \in C}} \frac{1}{e_{i}(a,p)}} & \parbox[t]{.15\textwidth}{if $i$ is supported $\forall i \in C$}\\
		0 & \text{otherwise} \\
	       \end{dcases}
    \label{eq:pennycook}
\end{equation}

\noindent where \rev{C} is the set of supported CPU models and compilers, \textit{a }is the algorithm, and \({e_{i}}(a, p)\) is the normalized run time of the algorithm on CPU {\textit{i}} for problem \textit{p}. To assess if the portability is affected by the problem size we computed and analyzed the metric for the various batch sizes.

\section{Results}
\label{section:results}
\begin{figure}[ht]
   \includegraphics [width=1\columnwidth] {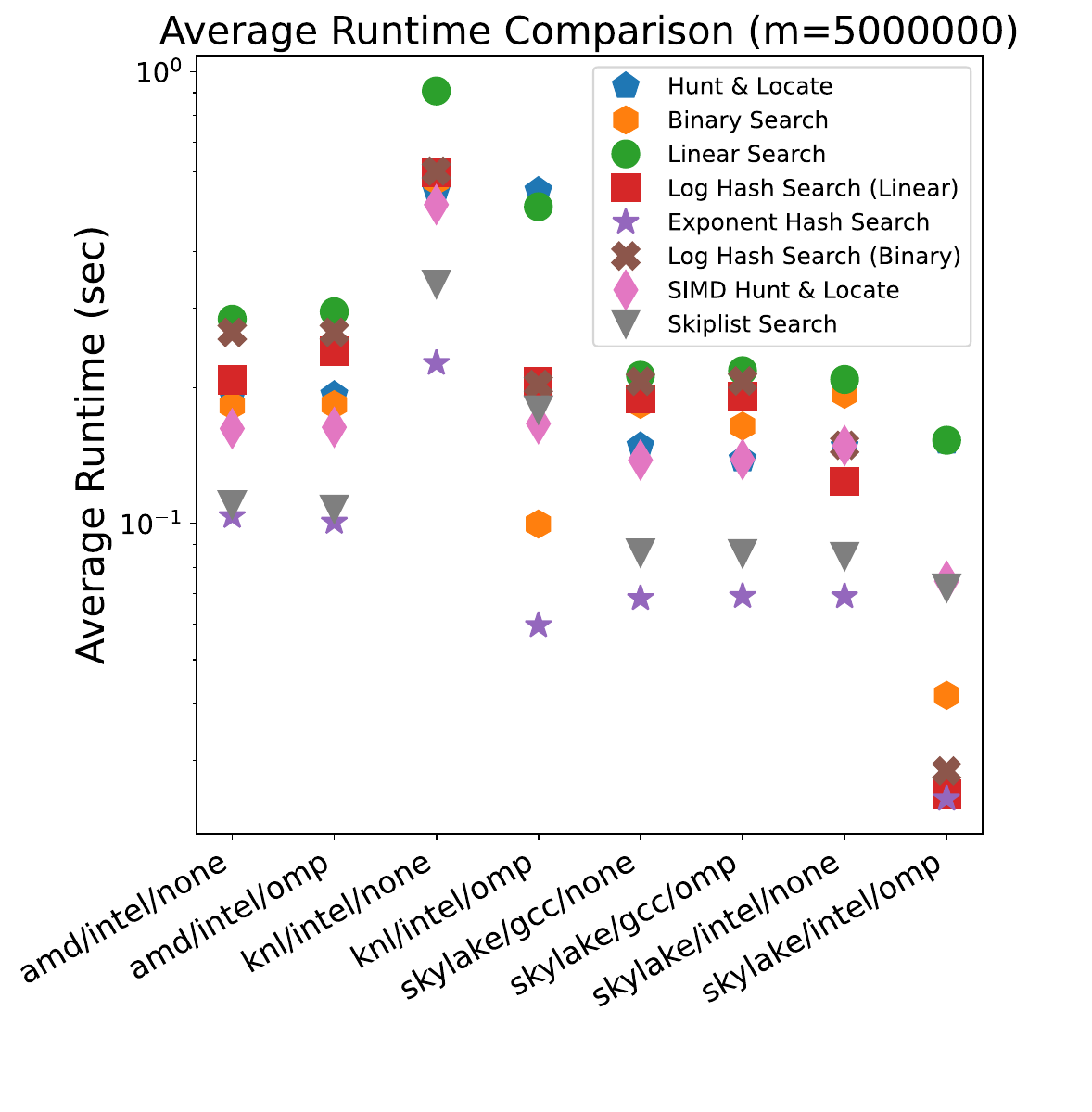}
   \centering
   \caption{Average run time comparison between configurations, labelled \textit{cpu/compiler/vectorization}}
   \label{fig:runtime_comparison}
\end{figure}

\begin{figure}[ht]
   \includegraphics [width=1\columnwidth] {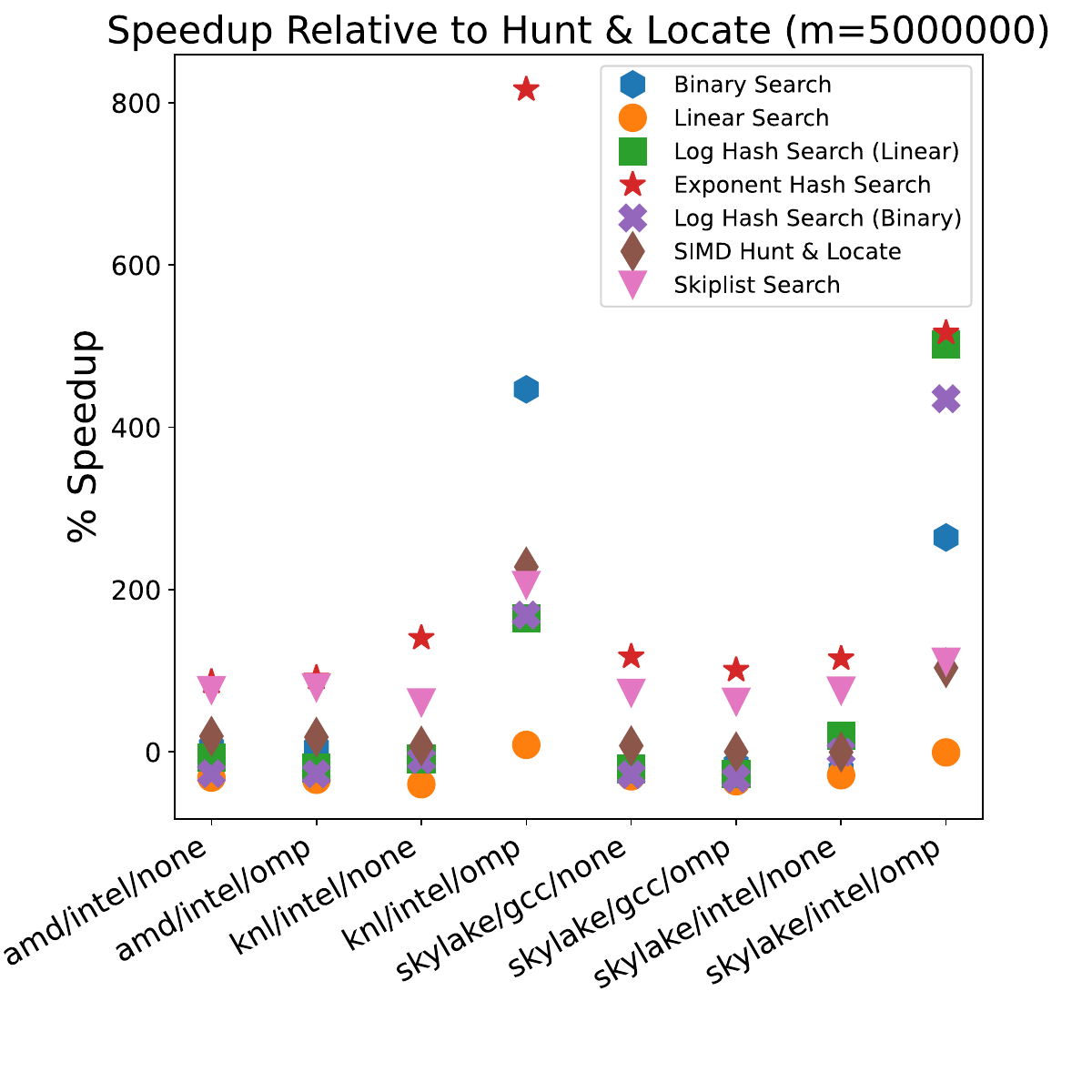}
   \centering
   \caption{Speedup relative to Hunt \& Locate}
   \label{fig:speedup_comparison}
\end{figure}

To demonstrate the performance of our algorithms, we measured their running time with $m=5000000$ (targets array size), $n=110$ (data array size), averaged over 20 trials, using configurations of both GCC and Intel compilers on several CPU models, with vectorization both enabled and disabled using compiler flags. The shorthand labels for the CPU models used in the performance plots is shown in Table~\ref{table:cpu_model_info}. Figure~\ref{fig:runtime_comparison} shows an average run time for each configuration of CPU and compiler. GCC was omitted except for a single configuration because there were no significant performance increases observed from vectorization. A refinement of the performance comparisons is shown in Fig.~\ref{fig:speedup_comparison} where each search algorithm tested is compared to the baseline search algorithm, \textit{Hunt and Locate}. For each CPU model, the performance of GCC was comparable to unvectorized Intel regardless of whether vectorization was enabled.

\begin{table}
\resizebox{\columnwidth}{!}{%
\begin{tabular}{|c|c|c|}
        \hline
        \textbf{Label}  & \textbf{CPU Model}                & \textbf{SIMD width} \\
        \hline
        amd             & AMD EPYC 7551 32-Core             & 256 \\
        knl             & Intel(R) Xeon Phi(TM) CPU 7250    & 512 \\
        skylake         & Intel(R) Xeon(R) Gold 6152 CPU    & 512  \\
        \hline
    \end{tabular}
    }
    \caption{CPU model legend for runtime graphs}
    \label{table:cpu_model_info}
\end{table}

\subsection{Array Sizes}
For testing we chose to use a large targets array size, $m$, relative to the data array size, $n$, because it is the most relevant use case for table-lookups in general and for our specific implementation in EOSPAC. The hash-based search algorithms perform best with these parameters. Hashtable generation time is $O(n)$, so when $m \gg n$, hashtable generation time becomes insignificant.

$m \gg n$ also has a significant impact on the viability of vectorization. As the outermost loop, which is over $Y$, of length $m$, must be the vectorized loop, the amount of work in its body varies by a factor of $n$. If $n$ is small, there is little penalty due to masking. When $n$ is large, the impact may be disproportionately large, sometimes even preventing vectorization from being viable at all.

\begin{figure}[ht]
   \includegraphics [width=1\columnwidth] {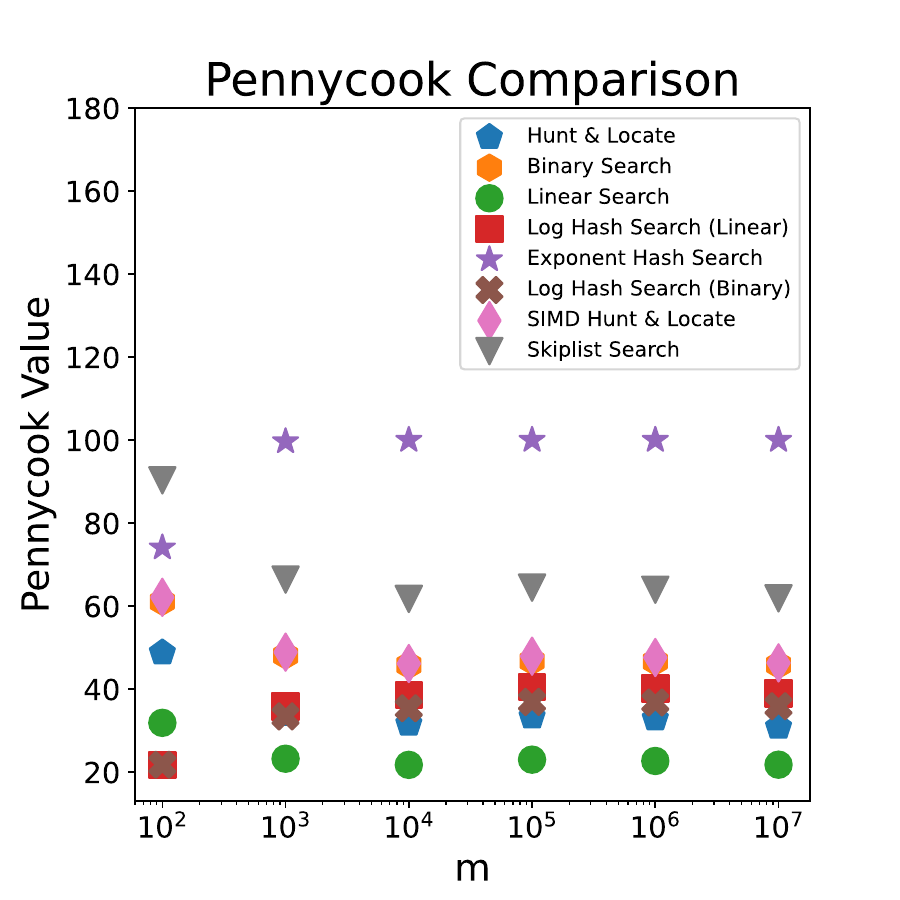}
   \centering
   \caption{The Pennycook performance portability}
   \label{fig:Pennycook Comparison}
\end{figure}

The effects of this can be seen in Fig.~\ref{fig:Pennycook Comparison}. As $m$ approaches $10^2$, the lead in performance of \textit{Exponent Hash Search} is lost to \textit{Skiplist Search}, due to its much cheaper setup cost ($O(n/8)$). The figure uses the Pennycook formula shown in Fig.~\ref{eq:pennycook}, which measures the performance portability across different configurations of CPU model and compiler.

\subsection{Data Array Distribution}
For hashtable generation, performance is primarily affected by the way the data array is distributed. In our tests, the data array roughly follows a logarithmic distribution with a value range of $3 \times 10^{-6}$ to $5.4 \times 10^4$. This is close to optimal for \textit{Exponent Hash Search} [Section~\ref{sect:exponent_hash_search}], which is faster the closer each value in $X$ is to being a unique power of 2. This is because the hash function extracts the power of 2 directly from the exponent bits of each target value. In this case, there are about 34 powers of 2 between the smallest and largest value, causing the number of indices between each hashtable entry to be $110/34=3.235$. If the data had a much narrower or wider range, \textit{Logarithm Hash Search} [Section~\ref{sect:log_hash_search}] would perform better, as it adjusts the base of its hash function to fit the range of values in the data array. 

\rev{\textit{Skiplist Search}, \textit{Binary Search} and \textit{Hunt and Locate} are less sensitive to data distribution, although they perform badly in the worst-case scenario where the target element is at either the beginning or end of the data array.}

\subsection{Search Call Frequency}
The number of search calls has a significant impact on performance for algorithms with a non-zero setup cost. In our experimental setup, only one call to search was made. When using algorithms that incur a setup cost for a table lookup library such as EOSPAC where many search calls are made, a more involved design decision must be made. For EOSPAC, the arrays that are to be searched over are loaded only once at the beginning of the user program. This allows the search setup to also be performed only once. In situations where there is not a known set of data arrays to be searched over, a different approach must be taken, possibly introducing greater complexity than the search algorithm alone. In such cases, algorithms with zero setup cost may be a better choice.

\subsection{Vectorization Considerations}
Directing a compiler to effectively vectorize code is challenging. We tried many different combinations of flags for both GCC and Intel. The flags that worked best for performance are documented in the Appendix~\ref{appendix}. While we were unable to get GCC to vectorize the experimental setup, we were successful in getting Intel to vectorize effectively. This is likely due to differences in the maturity of the vectorization implementations between the two compilers. 

\begin{figure}[ht]
   \includegraphics [width=1\columnwidth] {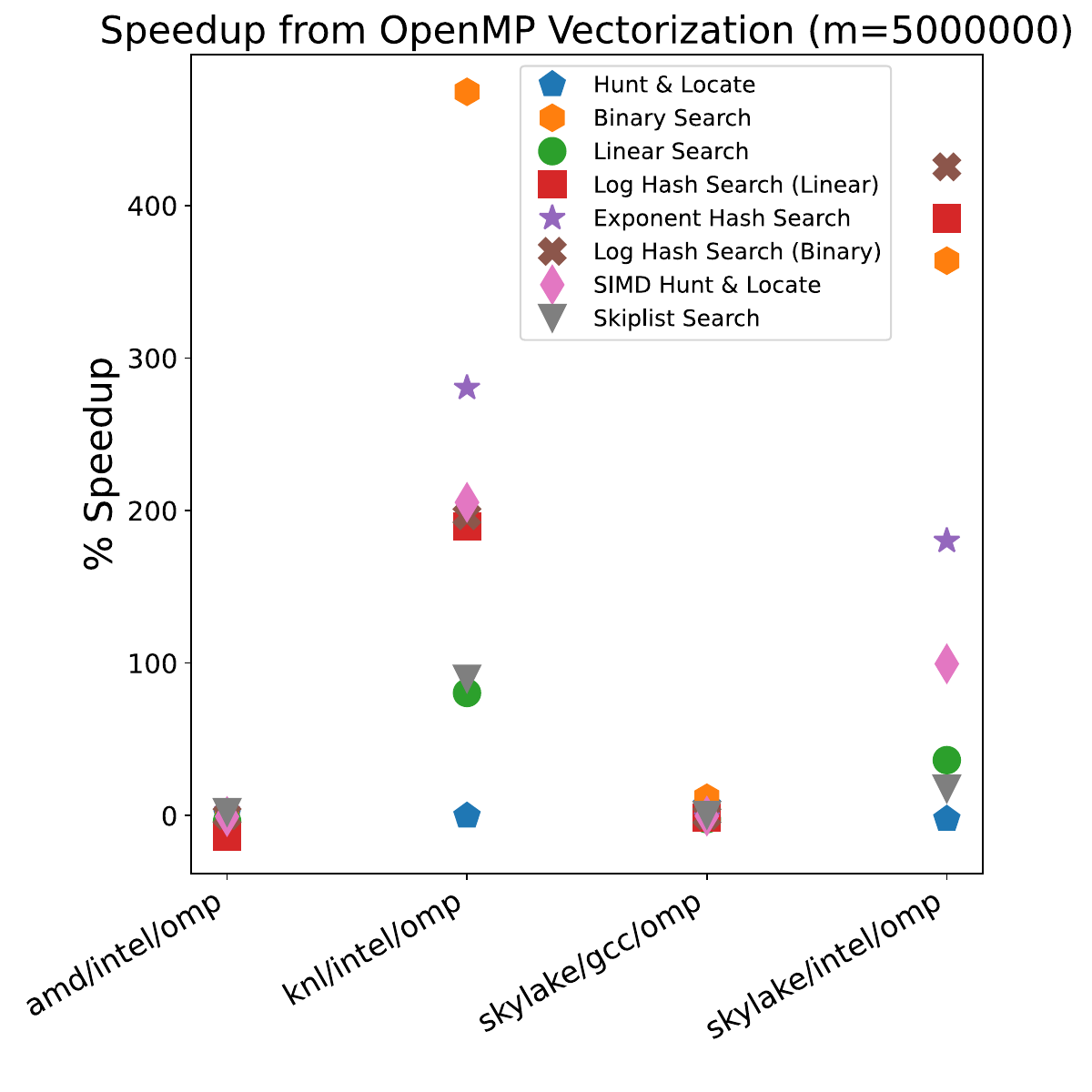}
   \centering
   \caption{Speedup from OpenMP vectorization relative to unvectorized}
   \label{fig:vectorization_speedup_comparison_omp}
\end{figure}

\begin{figure}[ht]
   \includegraphics [width=1\columnwidth] {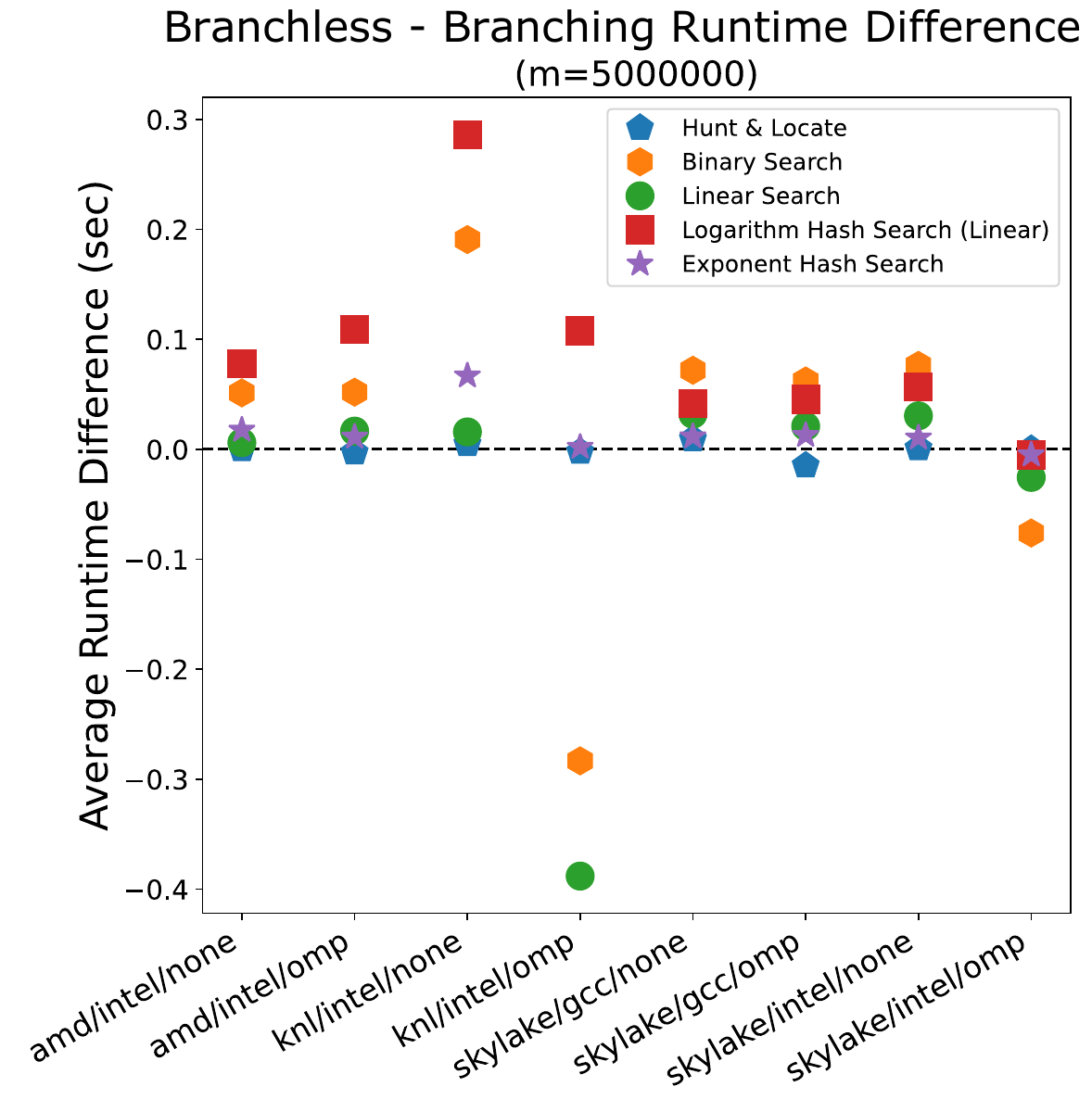}
   \centering
   \caption{Comparison of \rev{branching and branchless versions of algorithms. Points below the line show algorithms that were faster when branching was removed.}}
   \label{fig:optimization_performance}
\end{figure}

Vectorization had a significant impact on performance when it was enabled through the Intel compiler as shown in Fig.~\ref{fig:vectorization_speedup_comparison_omp}, with varying results depending on CPU model. The design of search algorithms must be constrained in particular ways to benefit from vectorization. For the compiler to generate SIMD instructions, memory access must be linear, and there can be no dependence between loop iterations. A linear control flow is also necessary for SIMD instructions to be generated. If the code's control flow is not linear, the compiler will compensate by generating masked SIMD instructions. This forces all computation of all possible branches to be performed, masking out the result of the branches not taken. The performance bottleneck then becomes the slowest branch. The Intel compiler was able to vectorize even when all branching was left in place, but performance suffered due to masking. To address this issue, branching was removed as much as possible so that fewer masked instructions were generated. However, in some cases, this caused a loss in performance when vectorization was disabled, due to the loss of branch prediction and other compiler optimizations relating to control flow. The resulting performance due to these effects is shown in Fig.~\ref{fig:optimization_performance}. This trade-off must be taken into account when integrating these algorithms into an application.

When vectorization is possible, \textit{Logarithm Hash Search} [Section~\ref{sect:log_hash_search}] typically outperforms \textit{Exponent Hash Search} [Section~\ref{sect:exponent_hash_search}] when the value range in the data array is narrow. When vectorization is not possible, \textit{Skiplist Search} [Section~\ref{sect:skiplist_search}] becomes a more viable alternative if the data is not distributed logarithmically. In general, we found that \textit{Exponent Hash Search} performs the best overall when vectorization is possible, as the hash function is composed of very cheap operations. \textit{Logarithm Hash Search}'s hash function uses the C library logarithm function, which depends on glibc to be decomposed into SIMD instructions. In our case, GCC was unable to perform this transformation, but with newer versions of glibc and GCC it may be possible to do so. The Intel compiler was able to generate SIMD instructions for logarithm, but they are still more computationally expensive than those of \textit{Exponent Hash Search}.

The algorithms that responded best to vectorization were \textit{Binary Search} [Section~\ref{sect:binary_search}], the hash-based searches [Section~\ref{sect:exponent_hash_search}], [Section~\ref{sect:log_hash_search}], and \textit{Hunt and Locate} [Section~\ref{sect:hunt_and_locate}]. These successes are correlated with consistent workload and linear data access pattern between searches. Since the completion of each concurrent batch of operations performed by SIMD vector instructions is constrained by the longest running operation in the batch, improvements to the operations' workload consistency will waste fewer computation cycles. A linear data access pattern is important because it leads to better cache performance.

\subsection{Search Array Distribution}
The search array for testing consisted of values sampled from a uniform random distribution. Approximately half of the values were outside of the bounds of the data array. This best represents the worst case of a typical problem, the best case being a sorted search array with no out-of-bounds values. A sorted search array would be much better suited for algorithms such as the original EOSPAC \textit{Hunt and Locate}, which uses the result of each search as the starting point for the next. However, this advantage may be superseded by vectorization, which cannot occur if there is a dependency between search iterations.

\subsection{Compiler Flags}
We found that the most effective method of vectorization for the Intel compiler was explicitly enabling OpenMP vectorization through the \ttsml{-qopenmp-simd} flag. This enables extra optimizations that are unavailable when auto-vectorizing, as well as extra available configuration (such as array alignment) that is not available through auto-vectorization alone. The flags that work best for auto-vectorization should also be enabled alongside the \ttsml{-qopenmp-simd} flag. See Appendix~\ref{appendix} for a list of these flags.

\subsection{Cache Performance}
\begin{figure}[ht]
   \includegraphics[width=1\columnwidth]{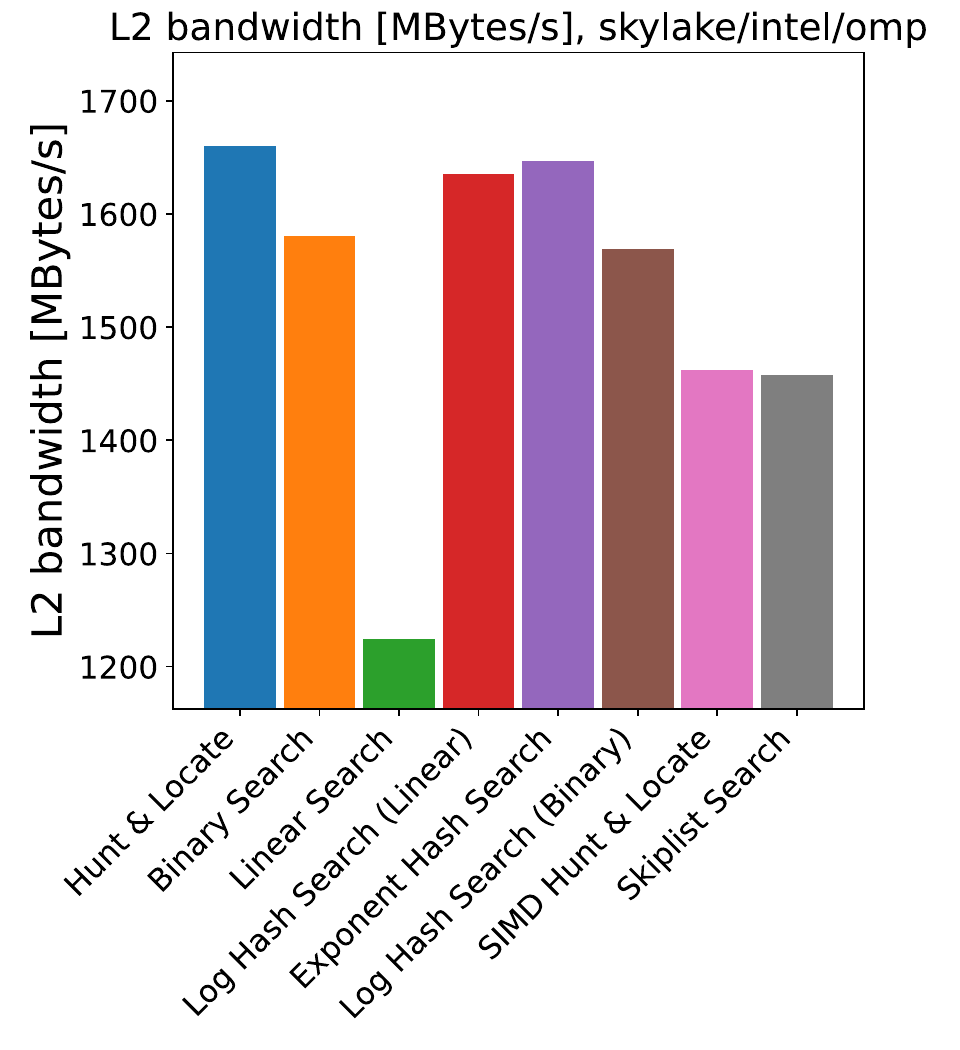}
   \caption{L2 Cache Bandwidth on Skylake Gold, Intel 20.0.2, OpenMP enabled}
   \label{fig:L2bandwidth}
\end{figure}

\begin{figure}[ht]
   \includegraphics[width=1\columnwidth]{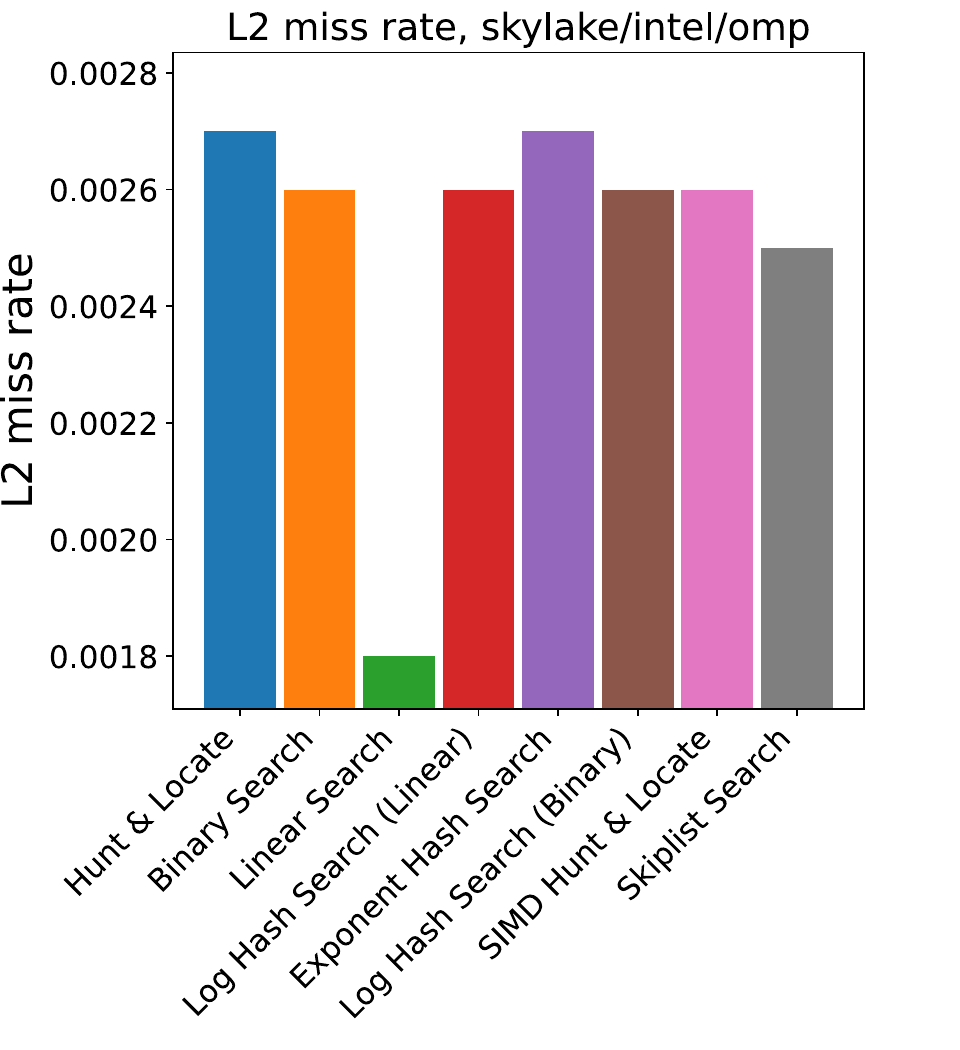}
   \caption{L2 Cache Miss Rate on Skylake Gold, Intel 20.0.2, OpenMP enabled}
   \label{fig:L2cachemiss}
\end{figure}

\begin{figure}[ht]
   \includegraphics[width=1\columnwidth]{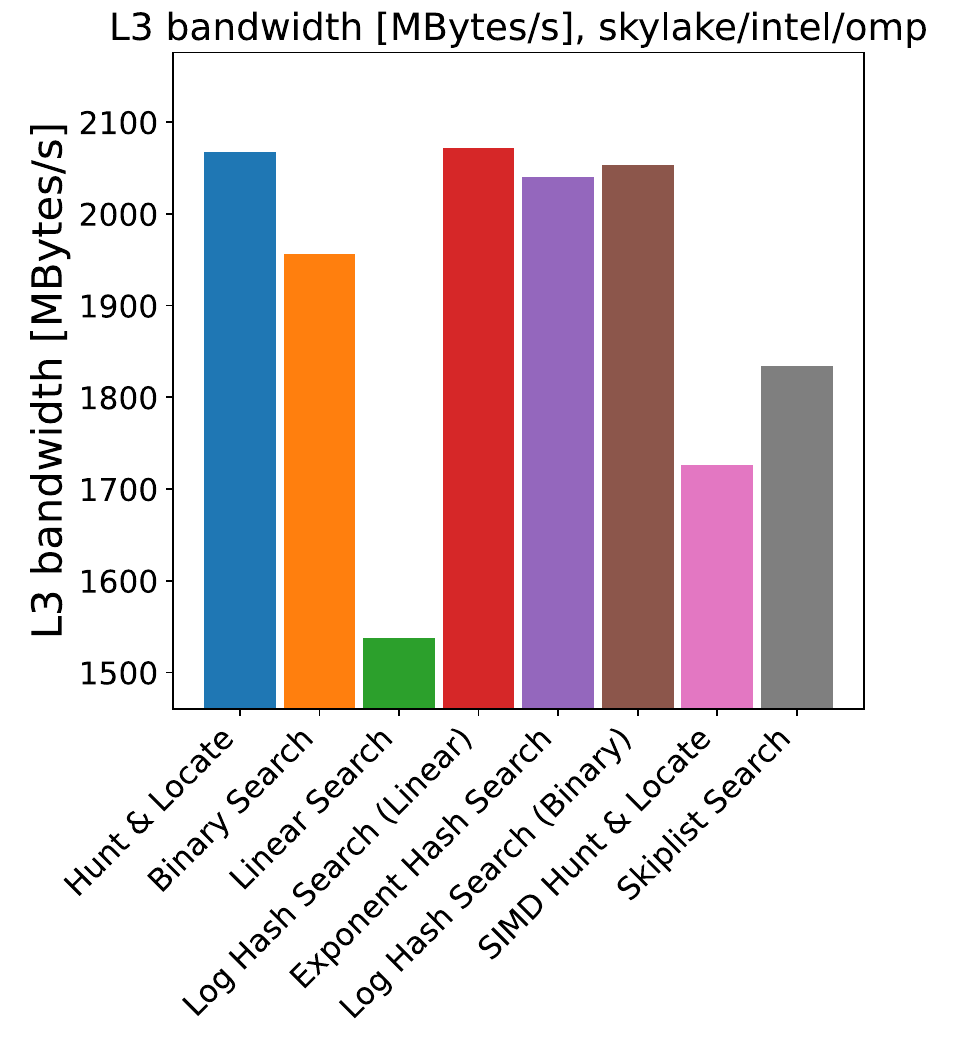}
   \caption{L3 Cache Bandwidth on Skylake Gold, Intel 20.0.2, OpenMP enabled}
   \label{fig:L3bandwidth}
\end{figure}

We used Likwid~\cite{likwid} to measure L1 Instruction cache TLB miss rate, L2 cache bandwidth, L2 cache miss rate, L3 cache bandwidth, and L3 cache miss rate. These values are useful for determining the cache performance of each algorithm. The following plots, Fig.~\ref{fig:L2bandwidth}, \ref{fig:L2cachemiss} and \ref{fig:L3bandwidth}, present these results. \rev{For these tests we used a search array size $m=5000000$.}

The cache performance of \textit{Linear Search} serves as an optimal baseline. \textit{Skiplist Search} [Section~\ref{sect:skiplist_search}] has good cache performance because the skiplist array is exactly the size of a single cache line. The hash-based searches do not perform as well, because of their hashtables' unpredictable data access patterns. \rev{However, \textit{Logarithm Hash Search} [Section~\ref{sect:log_hash_search}] often performs fewer searches, so it has slightly better cache performance than \textit{Exponential Hash Search} [Section~\ref{sect:exponent_hash_search}]}.

\section{Conclusion}

Several different search algorithms were analyzed in order to determine how best to improve the overall performance of search within the EOSPAC software package. \rev{Due to implementation constraints within the software that uses EOSPAC, we investigated performance gains of vectorizing search as opposed to using other parallelization techniques like threading or MPI\footnote{The Message Passing Interface (MPI) is an open library standard for distributed memory parallelization.}.} When compared to the original \textit{Hunt and Locate} algorithm, we saw a speedup by as much as 800\% with a search array of size 5000000 using the vectorized (e.g., OpenMP SIMD) \textit{Exponent Hash Search}. It was also shown that explicitly combining OpenMP vectorization and auto-vectorization compiler flags provided further performance enhancements beyond that of auto-vectorization alone by as much as 480\%. Even when vectorization is not possible, \textit{Exponent Hash Search} outperforms \textit{Hunt and Locate} by more than 100\%. The most consistent search algorithm performance improvement, according to the Pennycook metric, is the \textit{Exponential Hash Search} algorithm for search array sizes greater than 1000. The \textit{Skiplist Search} algorithm demonstrated the best performance under this threshold, due to its much smaller setup cost.

While cache utilization is important, it is not the only factor for vectorization and the resulting algorithm performance. We note that the worst performing algorithm (\textit{Linear Search}) exhibited the best cache performance. Good performance is dependent on a \rev{complex set} of factors, including the effects of hardware and compilers.

Several search algorithms have been investigated and the performance improvements of OpenMP vectorization have been determined to be beneficial as long as vectorization is adequately implemented by the chosen compiler.

\clearpage

\begin{acknowledgment}

This collaboration was made possible thanks to the Parallel Computing Summer Research Internship program, which is run out of the Information Science and Technology Institute, and the Equation of State Project, which is run out of the Advanced Simulation and Computing Program (ASC).
Additional technical support and collaboration was also provided by the members of the ASC Beyond Moore's Law Inexact Computing Project at Los Alamos National Laboratory.
This work was funded by the U.S. Department of Energy through the Los Alamos National Laboratory. Los Alamos National Laboratory is operated by Triad National Security, LLC, for the National Nuclear Security Administration of U.S. Department of Energy (Contract No. 89233218CNA000001).
The United States Government retains, and by accepting the article for publication, the publisher acknowledges that the United States Government retains, a non-exclusive, paid-up, irrevocable, worldwide license to publish or reproduce the published form of this work, or allow others to do so, for United States Government purposes.

\end{acknowledgment}

\appendix
\section{Appendix}
\label{appendix}
C code for the algorithms explored in this paper can be found at 
\newline {\small  \href{https://github.com/lanl/eospac/tree/main/simd\_search}{https://github.com/lanl/eospac/tree/main/simd\_search}}

\subsection{Compiler flags for vectorization}
\subsubsection{GCC}
{\fontsize{8}{11} \texttt{-fstrict-aliasing -ftree-vectorize -march=native -mtune=native -fopt-info-vec-all=gcc\_optrprt -fopenmp-simd -O3}}

\subsubsection{Intel}
{\fontsize{8}{11} \texttt{-ansi-alias -fp-model:precise -xHOST -vecabi=cmdtarget -qopt-zmm-usage=high -qopt-report=5 -qopt-report-phase=openmp,loop,vec -qopenmp-simd -march=native -O3}}

\subsection{Compiler flags for no vectorization}
\subsubsection{GCC}
{\fontsize{8}{11} \texttt{-fno-tree-vectorize -march=native -O3}}

\subsubsection{Intel}
{\fontsize{8}{11} \texttt{-no-vec -qno-openmp-simd -march=native -O3}}

\subsection{Preprocessor OpenMP Pragma}
This pragma was used to enable OpenMP SIMD functionality on the search loops in the algorithms tested:\\
{\fontsize{8}{11} \texttt{\#pragma omp simd aligned(x, y, jlower)}}\\
{\fontsize{8}{11} \texttt{jlower}} refers to the output lower-bounds array.

\bibliographystyle{asmems4}
\bibliography{EOSPAC-PCSRI_2020}
\end{document}